\input harvmac
\noblackbox
\newcount\figno
\figno=0
\def\fig#1#2#3{
\par\begingroup\parindent=0pt\leftskip=1cm\rightskip=1cm\parindent=0pt
\baselineskip=11pt
\global\advance\figno by 1
\midinsert
\epsfxsize=#3
\centerline{\epsfbox{#2}}
\vskip 12pt
\centerline{{\bf Figure \the\figno:} #1}\par
\endinsert\endgroup\par}
\def\figlabel#1{\xdef#1{\the\figno}}

\font\cmss=cmss10
\font\cmsss=cmss10 at 7pt
\def\rlx{\relax\leavevmode}
\def\inbar{\vrule height1.5ex width.4pt depth0pt}
\def\IC{\relax\,\hbox{$\inbar\kern-.3em{\rm C}$}}
\def\IN{\relax{\rm I\kern-.18em N}}
\def\IP{\relax{\rm I\kern-.18em P}}
\def\ZZ{\rlx\leavevmode\ifmmode\mathchoice{\hbox{\cmss Z\kern-.4em Z}}
 {\hbox{\cmss Z\kern-.4em Z}}{\lower.9pt\hbox{\cmsss Z\kern-.36em Z}}
 {\lower1.2pt\hbox{\cmsss Z\kern-.36em Z}}\else{\cmss Z\kern-.4em Z}\fi}
\def\narrowplus{\kern -.04truein + \kern -.03truein}
\def\narrowminus{- \kern -.04truein}
\def\narrowminussub{\kern -.02truein - \kern -.01truein}

\def\doubref#1#2{\refs{{#1},{#2}}}

\font\cmss=cmss10
\font\cmsss=cmss10 at 7pt
\def\rlx{\relax\leavevmode}

\def\inbar{\vrule height1.5ex width.4pt depth0pt}
\def\IB{\relax{\rm I\kern-.18em B}}
\def\IC{\relax\,\hbox{$\inbar\kern-.3em{\rm C}$}}
\def\ID{\relax{\rm I\kern-.18em D}}
\def\IE{\relax{\rm I\kern-.18em E}}
\def\IF{\relax{\rm I\kern-.18em F}}
\def\IG{\relax\,\hbox{$\inbar\kern-.3em{\rm G}$}}
\def\IH{\relax{\rm I\kern-.18em H}}
\def\II{\relax{\rm I\kern-.18em I}}
\def\IK{\relax{\rm I\kern-.18em K}}
\def\IL{\relax{\rm I\kern-.18em L}}
\def\IM{\relax{\rm I\kern-.18em M}}
\def\IN{\relax{\rm I\kern-.18em N}}
\def\IO{\relax\,\hbox{$\inbar\kern-.3em{\rm O}$}}
\def\IP{\relax{\rm I\kern-.18em P}}
\def\IQ{\relax\,\hbox{$\inbar\kern-.3em{\rm Q}$}}
\def\IR{\relax{\rm I\kern-.18em R}}
\def\ZZ{\rlx\leavevmode
             \ifmmode\mathchoice
                    {\hbox{\cmss Z\kern-.4em Z}}
                    {\hbox{\cmss Z\kern-.4em Z}}
                    {\lower.9pt\hbox{\cmsss Z\kern-.36em Z}}
                    {\lower1.2pt\hbox{\cmsss Z\kern-.36em Z}}
               \else{\cmss Z\kern-.4em Z}\fi}

\def\notin{\ \hbox{{$\in$}\kern-.51em\hbox{/}}}

\def\e{\epsilon}     
     
\def\Si{\Sigma}   
 \def\cB{{\cal B}} \def\cC{{\cal C}}
\def\cD{{\cal D}}

  \def\cL{{\cal L}}

\def\cN{{\cal N}} \def\cO{{\cal O}} 

 \def\cS{{\cal S}}

\def\lra{\longrightarrow}
\def\lolra{\longleftrightarrow}

\def\hbar{\bar h}

\font\large=cmr10 scaled \magstep3
\newif\ifnref

\def\doubref#1#2{\refs{{#1},{#2}}}

\nreffalse
%

\def\doubref#1#2{\refs{{#1},{#2}}}

\lref\rew{E.\ Witten, {\it Nonperturbative Superpotentials in String
       Theory}, Nucl.\ Phys.\ {\bf B474}(1996)343, hep-th/9604030}
\lref\rewone{E.\ Witten, {\it String Theory Dynamics in Various Dimensions},
     Nucl.Phys. {\bf B443}(1995)85, hep-th/9503124}
\lref\rht{C.\ Hull and P.\ K.\ Townsend, {\it Enhanced Gauge Symmetries
    in Superstring Theory}, Nucl.Phys. {\bf B451}(1995)525, 
    hep-th/9505073} 
\lref\rjh{J.\ Schwarz, {\it The power of M theory},
  Phys.\ Lett.\ {\bf B367} (1996) 97, hep--th/9510086} 
\lref\rhw{P.\ Horava and E.\ Witten,
  {\it Heterotic and type I string dynamics from eleven dimensions},
  Nucl.\ Phys.\ {\bf B460} (1996) 506, hep--th/9510209;
  {\it Eleven-dimensional supergravity on a manifold with boundary},
  hep--th/9603142}
\lref\rcvone{C.\ Vafa, {\it Evidence for F--theory},
  Nucl.\ Phys.\ {\bf B469} (1996) 403, hep--th/9602022}
\lref\rmv{D.\ R.\ Morrison and C.\ Vafa, {\it Compactifications of 
   F--theory on Calabi--Yau threefolds {\rm I,II}},
  hep--th/9602114, hep--th/9603161}
\lref\rkv{S.\ Kachru and C.\ Vafa, {\it Exact Results for N=2
         Compactifications of Heterotic Strings},
      Nucl.\ Phys.\ {\bf B450}(1995)69, hep-th/9505105}
\lref\rgmp{B.\ R.\ Greene, D.\ R.\ Morrison and R.\ Plesser, {\it Mirror 
  Symmetry in Higher Dimensions}, Commun.\ Math.\ Phys.\ {\bf 173}(1995)559, 
     hep-th/9402119}  
\lref\rsk{S.\ Katz, {\it Rational curves on Calabi-Yau manifolds: 
       verifying predictions of Mirror Symmetry}, alg-geom/9301006} 
\lref\ras{A.\ Sen, {\it Duality and Orbifolds}, hep-th/9604070}
\lref\rkr{A.\ Kumar and K.\ Ray, {\it Compactifications of M-Theory to 
    Two Dimensions}, hep-th/9604164}
\lref\rewmarch{E.\ Witten, {\it Phase Transitions in M-Theory and 
     F-Theory}, hep-th/9603150} 
\lref\rsvw{S.\ Sethi, C.\ Vafa and E.\ Witten, {\it Constraints on
     Low-Dimensional String Compactifications}, hep-th/9606122}
\lref\rbs{I.\ Brunner and R.\ Schimmrigk, {\it F-Theory on Calabi-Yau 
 Fourfolds}, Phys.Lett.\ {\bf B387}(1996)750, hep-th/9606148} 
\lref\rdgw{R.\ Donagi, A.\ Grassi and E.\ Witten, {\it Nonperturbative
     Superpotential with $E_8$ Symmetry}, hep-th/9607091}
\lref\rgm{R.\ Gopakumar and S.\ Mukhi, {\it Orbifold and Orientifold 
  Compactifications of F-Theory and M-Theory to Six and Four Dimensions},
  hep-th/9607057} 
\lref\rsr{S.\ Roy, {\it An Orbifold and an Orientifold of Type IIB 
     Theory on K3$\times $K3}, hep-th/9607157}  
\lref\rbfpss{M.\ Bianchi, S.\ Ferrara, G.\ Pradisi, A.\ Sagnotti and 
     Ya.\ S.\ Stanev, {\it Twelve-Dimensional Aspects of Four-Dimensional 
      N=1 Type I Vacua}, hep-th/9607105}
\lref\rbbmooy{K.\ Becker, M.\ Becker, D.\ R.\ Morrison, H.\ Ooguri, 
 Y.\ Oz and Z.\ Yin, {\it Supersymmetric Cycles in Exceptional Holonomy 
 Manifolds and Calabi-Yau 4-Folds}, hep-th/9608116}
\lref\rks{S.\ Kachru and E.\ Silverstein, {\it Singularities, Gauge
  Dynamics, and Nonperturbative Superpotentials in String Theory},
  hep-th/9608194}
\lref\rcdls{P.\ Candelas, A.\ M.\ Dale, C.\ A.\ L\"utken and 
     R.\ Schimmrigk,
  {\it Complete intersection Calabi--Yau manifolds},
  Nucl.\ Phys.\ {\bf B298} (1988) 493}
\lref\rcgh{P.\ Candelas, P.\ Green and T.\ H\"ubsch, {\it Rolling 
     among Calabi-Yau Vacua}, Nucl.\ Phys.\ {\bf B330}(1990)49}  
\lref\rls{M.\ Lynker and R.\ Schimmrigk, {\it Conifold Transitions 
 and Aspects of Duality}, to appear in Nucl.\ Phys.\ {\bf B}, 
 hep-th/9511058}
\lref\rcggk{T.M.Chiang, B.R.Greene, M.Gross and Y.Kanter, 
  {\it Black Hole Condensation and the Web of Calabi-Yau Manifolds}, 
   hep-th/9511204}
\lref\racjm{A.\ C.\ Avram, P.\ Candelas, D.\ Jancic and M.\ Mandelberg, 
   {\it On the Connectedness of the Moduli Space of Calabi-Yau 
   Manifolds}, hep-th/9511230}
\lref\randys{A.\ Strominger,
  {\it Massless black holes and conifolds in string theory},
  Nucl.\ Phys.\ {\bf B451} (1995) 96, hep--th/9504090} 
\lref\rgms{B.\ R.\ Greene, D.\ R.\ Morrison and A.\ Strominger,
  {\it Black hole condensation and the unification of string vacua},
  Nucl.\ Phys.\ {\bf B451} (1995) 109, hep--th/9504145}
\lref\rcvtwo{C.\ Vafa, {\it A Stringy Test of the Fate of the Conifold}, 
      Nucl.Phys.\ {\bf B447}(1995)252}  
\lref\rgv{D.\ Ghoshal and C.Vafa, {it c=1 String as the Topological 
        Theory of the Conifold}, Nucl.Phys. {\bf B453}(1995)121, 
      hep-th/9506122}   
\lref\rafiq{G.\ Aldazabal, A.\ Font, L.E.\ Ib\'{a}\~{n}ez and 
      F.\ Quevedo, {\it Chains of N=2, D=4 Heterotic/Type II Duals}, 
         Nucl.Phys.\ {\bf 461}(1996)85} 
\lref\rdg{D.\ Ghoshal, {\it Notes of a singular Landau-Ginzburg family}, 
     Phys.\ Lett.\ {\bf B381}(1996)113, hep-th/9510233}
\lref\rov{H.\ Ooguri and C.\ Vafa, {\it Two-Dimensional Black Hole and 
     Singularities of CY Manifolds}, hep-th/9511164} 
\lref\rkm{A.\ Klemm and P.\ Mayr, {\it Strong Coupling Singularities
   and Non-Abelian Gauge Symmetries in N=2 String Theory},
    hep-th/9601014}
\lref\rcf{P.\ Candelas and A.\ Font, {\it Duality between the Webs of 
   Heterotic and Type II Vacua}, hep-th/9603170}
\lref\rbkk{P.\ Berglund, S.\ Katz, A.\ Klemm and P.\ Mayr, 
  {\it New Higgs Transitions between Dual N=2 String Models}, 
    hep-th/9605154}
\lref\rcpr{P.\ Candelas, E.\ Perevalov and G.\ Rajesh, {\it F-Theory 
   Duals of Nonperturbative Heterotic $E_8\times E_8$ Vacua in Six 
   Dimensions}, hep-th/9606133}  
\lref\rafiu{G.\ Aldazabal, A.\ Font,L.E.\ Ib\'{a}\~{n}ez and 
     A.M.Uranga, {\it New Branches of String Compactifications and 
       their F-Theory Dual}, hep-th/9607121}  
\lref\rgjns{E.\ Gava, T.\ Jayaraman, K.\ S.\ Narain and M.\ H.\ Sarmadi, 
    {\it D-Branes and the Conifold Singularity}, hep-th/9607131} 
\lref\rkmv{A.\ Klemm, P.\ Mayr and C.\ Vafa, {\it BPS States of Exceptional 
   Non-Critical Strings}, hep-th/9607139}  
\lref\rgmv{B.\ R.\ Greene, D.\ R.\ Morrison and C.\ Vafa, 
       {\it Geometric Realization of Confinement}, hep-th/9608039}
\lref\rovtwo{H.\ Ooguri and C.\ Vafa, {\it Summing up D--Instantons},
      hep-th/9608079}
\lref\rms{D.\ R.\ Morrison and N.\ Seiberg, {\it Extremal Transitions and 
        Five-Dimensional Supersymmetric Field Theories}, hep-th/9609070} 
\lref\rdkv{M.\ Douglas, S.\ Katz and C.\ Vafa, {\it Small Instantons, 
     Del Prezzo Surfaces and Type I' Theory}, hep-th/9609071} 
\lref\rweighted{P.\ Candelas, M.\ Lynker and R.\ Schimmrigk, {\it 
       Calabi-Yau Manifold in weighted $\IP_4$}, 
          Nucl.Phys.\ {\bf B341}(1990)383\semi
       A.\ Klemm and R.\ Schimmrigk, {\it Landau-Ginzburg String Vacua}, 
         Nucl.Phys.\ {\bf B411}(1994)559, hep-th/9204060\semi
       M.\ Kreuzer and H.\ Skarke, {\it No Mirror Symmetry among 
            Landau-Ginzburg Vacua}, Nucl.Phys. {\bf B338}(1992), 
          hep-th/9205004} 
\lref\rklm{A.\ Klemm, W.\ Lerche and P.\ Mayr, {\it K3 Fibrations},
       Phys.Lett. {\bf B357}(1995)313, hep-th/9506112 }
\lref\rhly{S.\ Hosono, B.\ Lian and S.-T.\ Yau, {\it Calabi-Yau Varieties
      and Pencils of K3 Surfaces}, alg-geom/9603020}
\lref\rhls{B.\ Hunt and R.\ Schimmrigk, {\it Heterotic Gauge Structure  
       of Type II K3 Fibrations}, Phys.\ Lett.\ {\bf B381}(1996)427, 
        hep-th/9512138\semi 
         B.\ Hunt, M.\ Lynker and R.\ Schimmrigk, {\it Heterotic/Type II 
       Duality in D=4 and String/String Duality},  
       hep-th/9609082} 
\lref\rjw{J.\ Werner, {\it Kleine Aufl\"osungen spezieller dreidimensionaler 
         Variet\"aten}, Bonner Mathematische Schriften, {\bf 186}, 1987}
\lref\rgh{P.\ S.\ Green and T.\ H\"ubsch, {\it Connecting Moduli Spaces of 
        Calabi-Yau Threefolds}, Commun.Math.Phys. {\bf 119}(1988)431}  
\lref\rakms{A.\ C.\ Avram, M.\ Kreuzer, M.\ Mandelberg and H.\ Skarke, 
        {\it Searching for K3 Fibrations}, hep-th/9610154}  
\lref\rlsone{M.\ Lynker and R.\ Schimmrigk, {\it Landau-Ginzburg Theories 
      as Orbifolds}, Phys.Lett.\ {\bf B249}(1990)237}
\lref\rrs{R.\ Schimmrigk, {\it Critical String Vacua from Noncritical 
                Manifolds}, Phys.Rev.Lett.\ {\bf 70}(1993)3688, hep-th/9210062;
           {\it Mirror Symmetry and String Vacua from a Special Class of 
            Fano Varieties}, Int.\ J.\ Mod.\ Phys.\ {\bf A11}(1996)3049, 
           hep-th/9405086} 
\lref\rpm{P.\ Mayr, {\it Mirror Symmetry, N=1 Superpotentials and Tensionless 
            Strings in Calabi-Yau Fourfolds}, hep-th/9610162} 
\lref\rvw{C.\ Vafa and E.\ Witten, {\it Dual String Pairs with N=1 and N=2 
     Supersymmetry om Four Dimensions}, hep-th/9507050} 
\lref\rbsw{R.\ Blumenhagen, R.\ Schimmrigk and A.\ Wi\ss kirchen, 
        {\it The (0,2) Exactly Solvable Structure of Chiral Rings, 
          Landau-Ginzburg Theories and Calabi-Yau Manifolds}, 
       Nucl.Phys.\ {\bf B461}(1996)460, hep-th/9510055; 
      {\it (0,2) Mirror Symmetry}, hep-th/9609167}   
\Title{\vbox{\hbox{hep--th/9610195}
             \hbox{BONN--TH--96--12}
             \hbox{HUB--EP--96/54}}}
{\bf {\large Unification of M- and F-Theory Calabi-Yau Fourfold Vacua }}
\centerline{Ilka Brunner\footnote{$^1$}{e--mail:\
                                       brunner@qft1.physik.hu-berlin.de},
            Monika Lynker\footnote{$^2$}{e--mail:\ mlynker@siggy.iusb.edu}
                 \ \ and \
            Rolf Schimmrigk\footnote{$^3$}{e--mail:\
                                          netah@avzw02.physik.uni--bonn.de}
            }
\bigskip
\centerline{${}^1$\it Institute for Physics, Humboldt University,
                      10115 Berlin}
\smallskip
\centerline{${}^2$\it Department of Physics and Astronomy,
                       Indiana University South Bend,}
\centerline{\it South Bend, Indiana, 46634 }
\smallskip
\centerline{${}^3$\it Physics Institute, University of Bonn, 53115 Bonn}

\vskip .4truein 
\centerline{\bf Abstract}
\vskip .1in
\noindent
We consider splitting type phase transitions between Calabi-Yau
fourfolds. These transitions generalize previously known types
of conifold transitions between threefolds. Similar to conifold
configurations the singular varieties mediating the transitions 
between fourfolds connect moduli spaces of different dimensions, 
describing ground states in M- and F-theory with different numbers 
of massless modes as well as different numbers of cycles to wrap 
various p-branes around. The web of Calabi-Yau fourfolds obtained  
in this way contains the class of all complete intersection manifolds 
embedded in products of ordinary projective spaces, but extends also 
to weighted configurations. It follows from this that for some of 
the fourfold transitions vacua with vanishing superpotential are 
connected to ground states with nonzero superpotential.

\Date{10/96}

\vfill\eject
\newsec{Introduction}
     
\noindent                                         
It has been a longstanding problem to formulate a dynamics on the 
collective moduli space of string
theory which would allow to determine the physical ground
state of the string. A first step in this direction would be to 
have a criterion which distinguishes
between different vacua in an intrinsic manner. Such a
criterion has recently been found by Witten \rew\ in the context 
of M-- and F--theory \refs{\rewone\rht\rjh\rhw\rcvone{--}\rmv}.
Compactifications of these theories to three and four dimensions involve 
eight-dimensional manifolds, in particular Calabi-Yau fourfolds. 
This observation has sparked considerable interest in the previously 
little investigated class of K\"ahler fourfolds with vanishing 
first Chern class\footnote{$^{\diamond}$}{Two notable exceptions are refs. 
                      \doubref\rgmp\rsk.}.
Attention so far has focused mostly on orbifolds 
\refs{\ras\rkr\rgm{--}\rsr} and more general complete intersection 
spaces \refs{\rewmarch\rsvw\rbs\rdgw\rbfpss\rbbmooy\rks{--}\rpm}. 
Witten's observation shows that different
ground states of F-- and M--theory lead to non-vanishing (vanishing)  
superpotential because of the (non)existence of certain types
of divisors in the internal Calabi--Yau space.

The natural question then arises whether the moduli space of Calabi-Yau 
fourfolds is connected so that M- and F-theory can conceivably 
`tunnel' between these different types of ground states. 
A simple argument shows that this is to be expected, at least for certain 
types of fourfolds. Special among four-dimensional Calabi-Yau 
manifolds CY$_4$ are fibered spaces for which the generic 
(quasi-)smooth fiber is a Calabi-Yau threefold CY$_3$. For such  
fibrations we can use the known conifold transitions between threefolds 
\refs{\rcdls\rcgh{--}\rls}, or more severe transitions described by 
operations on the toric data \doubref\rcggk\racjm, to induce a transition 
in the fourfold by degenerating the fibers pointwise. 
The Hodge numbers of the fibers change in this process and therefore we 
expect to be able to link in this way the moduli spaces of 
cohomologically distinct fourfolds. The connectedness 
of the collective moduli space of Calabi-Yau threefolds therefore 
immediately implies the connectedness of at least some regions of the 
moduli space of fourfolds. 

The class of CY$_3$ fibered fourfolds is further distinguished because 
even though in F-theory on such spaces we are considering N$=$1 
supersymmetric theories in D$=$4 we expect that for this type of 
manifolds many of the N$=$2 results carry over by a fiber-wise 
application via the adiabatic limit argument of \rvw\ or the twist 
map construction of \doubref\rhls\rbs. Using the duality results of 
\refs{\rewone\rht\rjh\rhw\rcvone{--}\rmv}\rkv\ then makes it clear in 
particular that many of the physical aspects of the threefold transitions 
\refs{\randys\rgms\rcvtwo\rgv\rafiq\rdg\rov\rkm\rcf\rbkk\rcpr\rafiu
      \rgjns\rkmv\rgmv\rovtwo\rms{--}\rdkv} 
will have F-theory and M-theory counterparts, perhaps by utilizing 
heterotic string models based on the (0,2) Calabi-Yau threefolds 
considered in \rbsw. 

In the present paper we will take the first steps in this direction by 
showing that large classes of Calabi-Yau fourfolds are connected. 
In Section 2 we generalize to fourfolds the determinantal conifold 
transition between ordinary projective complete intersection Calabi-Yau 
threefolds \rcdls\ and its weighted extension \rls.  Whereas in the 
case of threefolds the degenerations are rather mild, involving only 
conifold configurations with a finite number of 
nodes, the higher dimensional transitions we are going to describe 
proceed via singular varieties which involve degenerations for which 
the singular sets are two-dimensional, described in general by 
disconnected configurations of algebraic curves. Similar to the case 
of threefolds, however, the singular sets can be resolved in two 
different ways, either by deforming the degenerate variety along 
some complex modulus or by performing a small resolutions. Each 
of these ways to resolve the singularities leads to a (quasi-)smooth 
Calabi-Yau manifold with a different Hodge diamond. 
In Section 3 we describe a second splitting construction for fourfolds 
which is based on considerations of discriminantal varieties, discussed 
in the framework of ordinary projective complete intersection Calabi-Yau 
threefolds in \rgh.

The web of Calabi-Yau fourfolds obtained by these splitting constructions 
contains as a subset the class of all complete intersection manifolds 
embedded in products of ordinary projective spaces. This web is 
further extended 
by connecting it to the collective moduli space of weighted complete 
intersection spaces. All the constructions described in the present paper 
are independent of the fiber structure of the varieties involved and  
the spaces connected may or may not be fibered.
 
In the final Sections we apply splitting to define transitions 
between F\&M-vacua on fourfolds and show that direct splits can 
connect ground states with zero superpotential to those 
with non-vanishing superpotential. It turns out that it is the small 
resolution of certain configurations of singular curves contained in 
the degenerate varieties which mediate the transitions that can generate 
divisors with the needed properties to give rise to a non-vanishing 
superpotential.
This leads to the possibility that certain splitting type phase transitions 
between Calabi-Yau fourfolds lead to supersymmetry breaking. 

\newsec{Splitting Transitions between Fourfolds}
\noindent
Our focus in the following will be on complete intersection manifolds 
contained in configurations of the type 
\eqn\wcicys{
\matrix{\IP_{(k_1^1,\dots ,k_{n_1+1}^1)}\cr
               \IP_{(k_1^2,\dots ,k_{n_2+1}^2)}\cr \vdots \cr
               \IP_{(k_1^F,\dots ,k_{n_F+1}^F)}\cr}
\left [\matrix{d_1^1&d^1_2&\ldots &d_N^1\cr
               d_1^2&d_2^2&\ldots &d_N^2\cr
               \vdots&\vdots&\ddots &\vdots\cr
               d_1^F&d_2^F&\ldots &d_N^F\cr}\right]~~=~~X.
}
Such configurations describe the intersection of the zero locus of $N$
 polynomials embedded in a product of weighted projective spaces,
where $N= \left(\sum_{i=1}^F n_i -4\right)$ is the number of 
polynomials $p_a$ of F--degree $(d_a^1,\dots,d_a^F)$. Even though our 
considerations can be applied to general intersection spaces our 
main interest is in manifolds for which the first Chern class 
\eqn\chernone{
c_1(X) = \sum_{i=1}^F\left[\sum_{l=1}^{n_i+1} k_l^i -
       \sum_{a=1}^N d_a^i \right]h_i  
}
vanishes. Here we denote by $h_i, i=1,...,F$ the pullback of the generators 
of $H^2(\IP_{(k_1^i,\dots ,k_{n_1+1}^i)})$. 
Useful for the following will be the remaining Chern classes of weighted complete 
intersection Calabi-Yau fourfolds  
\eqn\cherning{\eqalign{
c_2(X) &= {1\over 2}\left[\sum_{a=1}^N \left(\sum_{i=1}^F d^i_a h_i\right)^2
          - \sum_{i=1}^F \sum_{r=1}^{n_i+1} (k_r^ih_i)^2 \right] \cr 
c_3(X) &= -{1\over 3}\left[\sum_{a=1}^N \left(\sum_{i=1}^F d^i_a h_i\right)^3
         - \sum_{i=1}^F \sum_{r=1}^{n_i+1} (k_r^ih_i)^3 \right] \cr
c_4(X) &= {1\over 4}\left[\sum_{a=1}^N \left(\sum_{i=1}^F d^i_a h_i\right)^4
              - \sum_{i=1}^F \sum_{r=1}^{n_i+1} (k_r^ih_i)^4
           + 2c_2^2 \right]. \cr
}}
It follows from the structure of $c_4$ that the Euler numbers of complete 
intersection fourfolds embedded in products of ordinary projective spaces 
is always positive.

\subsec{Determinantal Splitting Transitions}
There are several different types of transitions which one can 
construct between Calabi-Yau fourfolds. As the closest analog to the 
conifold transition one might consider the situation in which the 
fourfolds degenerate into varieties for which the singularities 
are again described by a finite number of singular points. 
Even in the case of threefolds it is not a simple matter 
in general to compute the resolution data and to check whether the 
resolved spaces are in fact again 
of Calabi-Yau type. A detailed investigation of the latter problem 
in the context of threefolds can be found in \rjw.
It is for this reason that the conifold transitions of 
splitting type are particularly simple - since they connect weighted 
complete intersection CYs per construction these global problems are 
resolved automatically. We therefore focus in the following on 
the four-dimensional generalization of the splitting construction.

\vskip .2truein
\noindent
{\sl $\IP_1$ Splits}\hfill
\break
\noindent
Consider the weighted complete intersection varieties of type \wcicys.
Introducing two vectors $u,v$ such that $(u^i+v^i)=d_1^i$
and denoting the remaining $(F \times (N-1))$--matrix by $M$, we 
write these spaces as $Y[(u+v)~~M]$. The simplest kind of transition 
is the $\IP_1$--split which is defined by  
\eqn\ponesplit{
X~=~Y[(u+v)~~M]~~\lolra ~~
\matrix{\IP_1\cr Y\hfill \cr} 
\left[\matrix{1&1&0\cr u&v&M\cr}\right]~=~X_{\rm split}.
}
The split variety of the rhs is described by the polynomials 
of the original manifold and two additional polynomials,  
which we can write as 
\eqn\splitpolly{\eqalign{
p_1 &=x_1Q(y_i) + x_2R(y_i) \cr
p_2 &=x_1S(y_i) + x_2T(y_i), \cr }}
where $Q(y_i), R(y_i)$ are of multi-degree $u$ and $S(y_i), T(y_i)$ 
are of degree $v$. 
In \splitpolly\ we collectively denote the coordinates of the space 
$Y$ by $y_i$ whereas the $x_i$ are the coordinates of the projective 
line $\IP_1$.

Insight into the precise relation of these two manifolds is obtained 
by comparing their Euler numbers, which can be obtained by Cherning. 
Since we are only interested in the local geometry of the transition 
we neglect for the moment possible orbifold 
singularities\footnote{$^{\dagger}$}{In general it is possible in weighted 
   manifolds for hypersurface singularities to sit on top of orbifold 
   singularities. In this case the following formulae have to be modified 
   in analogy to the analysis of ref. \rls\ for threefolds.} 
and suppose that the ambient space is a product of ordinary projective 
spaces $Y=\prod_{i=1}^F \IP_{n_i}$. 
We denote the K\"ahler form of the split factor $\IP_1$ by $H_0$ 
and the K\"ahler form of the $i^{th}$ factor by $H_i$. The Euler 
numbers are then obtained by integrating the top form over the ambient 
space. Because the generators $H_i$ of H$^2(\IP_{n_i})$ 
are normalized such that $\int_{\IP_{n_i}}H_i^{n_i}=1$ the coefficient of 
$H_0\prod_i H_i^{n_i}$ is precisely the Euler characteristic. 
Using the formulae \cherning\ one can then show that the difference 
between the Euler numbers of the two manifolds of the 
split \ponesplit\ are related by 
\eqn\eulrel{
\left(\chi(X_{\rm split}) - \chi(X)\right)\prod_{i=1}^F H_i^{n_i} 
= -3(u+v) u^2v^2 \prod_{a=2}^N \sum_{i=1}^F d^i_aH_i,}
where we have abused notation by writing the first two 
components $\cN_a$ of the normal bundle 
$\cN = \oplus_{a=0}^N \cN_a$ of the split manifolds as 
$c_1(\cN_0)=(H_0+u)$ and $c_1(\cN_1)=(H_0+v)$ with  
\eqn\redef{
u= \sum_{i=1}^F d^i_0H_i,~~~~~v=\sum_{i=1}^F d^i_1H_i.}

The result \eulrel\ shows that the split manifold describes a resolution 
of the determinantal 
variety in $Y[(u+v)~~M]$ defined by the original polynomials   
and the determinantal polynomial 
\eqn\detervar{
Y[(u+v)~~M] \ni X^{\sharp} =\{p_{\rm det} = QT - RS=0,~~p_a=0,~a=2,...,N\}
}
which can be viewed as the projection $\pi: X_{\rm split} \lra X^{\sharp}$ 
along the projective line $\IP_1$.  

To see this one notes that the hypersurface \detervar\ is singular 
on the locus 
\eqn\singloc{
\Si = Y[u~~u~~v~~v~~M]}
which describes (generically) a curve because $Y[(u+v)~~M]$ is 
four-dimensional. The Euler number of this curve $\Si$ however 
can be determined via Cherning to be  
\eqn\curveul{
\chi(\Si)\prod_{i=1}^FH_i^{n_i}
~=~ -(u+v)u^2v^2 \prod_{a=2}^N \sum_{i=1}^F d^i_aH_i~,}
hence we obtain the Euler number relation 
\eqn\eulrel{
\chi(X_{\rm split}) = \chi(X) + 3~\chi(\Si).}

We therefore see that we can smooth out the determinantal variety 
in two different ways: first by adding 
an appropriate deformation to 
$$
p_{\rm def}=p_{\rm det} + t\cdot p_{\rm trans} 
$$
which deforms the non-transverse determinantal polynomial into a 
transverse polynomial $p_{\rm def}$. In contrast to threefolds 
\refs{\rcdls\rcgh{--}\rls}, where the singular locus 
of the splitting transition is formed by a number of nodes, i.e.  
to a conifold configuration, for fourfolds the singular locus is 
an algebraic curve, i.e. a real two-dimensional surface with, 
in general, several components.
The important point however is that the singular set again 
admits a small resolution which,  
for fourfolds, involves the projective plane $\IP_2$ instead of 
the projective line $\IP_1$ of the threefold. Performing such a 
small resolution leads to the higher codimension split manifold.
Thus we arrive at the same singular space by degenerating two 
distinct manifolds in different ways 
$$ X \lra X^{\sharp} \longleftarrow X_{\rm split}.  
$$
Put differently, we can start from a determinantal variety and 
smooth out the singularities in two distinct ways 
$$ X \longleftarrow X^{\sharp} \longrightarrow X_{\rm split}.$$ 
Important for the general picture is the following generalization 
of the determinantal $\IP_1$ split.

\vskip .2truein 
\noindent
{\sl $\IP_n$ Splits} \hfill 
\break
\noindent
A similar discussion applies to the generalized $\IP_n$-split
\eqn\pnsplit{
X = Y\left[\sum_{a=1}^{n+1}u_a~~~M\right]~~\lolra ~~
\matrix{\IP_n\cr Y \hfill \cr}
\left[
      \matrix{1&1&\cdots &1 &0\cr u_1&u_2&\cdots &u_{n+1} &M\cr}
   \right]~=~X_{\rm split}
}
for which the Euler relation takes the form
\eqn\geneulrel{\eqalign{
(\chi(X_{\rm split}) &- \chi(X))\prod_{i=1}^F H_i^{n_i}~~~~~~~~~ \cr 
=&~~~{3\over 10}\left[ 4\sum_a u_a^3\sum_{b<c}u_bu_c
                  + 6\sum_a u_a^2 \sum_{b<c<d} u_bu_cu_d \right. \cr
 &~~~~~~~~~\left. + \sum_a u_a\left(4\sum_{b<c<d<e}u_bu_cu_du_e
                      - \sum_{{\rm \#}\{b,c,d,e\}>1} u_bu_cu_du_e\right)
        \right]\prod_{\beta} \xi_{\beta} \cr
}}
where
$$u_a = \sum_i d^i_a H_i $$
for $a=1,...,(n+1)$ and $\xi_{\beta}$ are the corresponding columns 
of matrix $M$ 
$$\xi_{\beta} = \sum_i m_{\beta}^i H_i.$$

Fourfold splits have a different local degeneration structure than   
threefold splits but they share certain features of these lower-dimensional 
counterparts. Most importantly the general $\IP_n$ splits immediately 
allow to connect all complete intersection Calabi-Yau fourfolds embedded 
in products of ordinary projective spaces 
\eqn\ordcicys{
\matrix{\IP_{n_1}\cr \IP_{n_2}\cr \vdots \cr \IP_{n_F}\cr}
\left [\matrix{d_1^1&d^1_2&\ldots &d_N^1\cr
               d_1^2&d_2^2&\ldots &d_N^2\cr
               \vdots&\vdots&\ddots &\vdots\cr
               d_1^F&d_2^F&\ldots &d_N^F\cr}\right]
}
to a particularly simple configuration, given by 
\eqn\motherordcicy{
\matrix{\IP_1\cr \IP_1\cr \IP_1\cr \IP_1\cr \IP_1\cr}
\left [\matrix{2\cr 2\cr 2\cr 2\cr 2\cr}\right]_{1440}
}
with Euler number $\chi=1440$ which can be determined via Cherning.
From Lefshetz' hyperplane theorem we know that $h^{(1,1)}=5$ and 
$h^{(2,1)}=0$. The dimension of  H$^{(3,1)}$ for this manifold can be 
determined by counting complex deformations with the result $h^{(3,1)}=227$.
From the Euler number we can then determine that final remaining 
Hodge number to obtain the complete Hodge diamond 
\eqn\mhodge{
\matrix{      &   &       &     &1       &     &      &    &   \cr
              &   &       &0    &        &0    &      &    &   \cr
              &   &0      &     &5       &     &0      &    &   \cr
              &0  &       &0    &        &0    &       &0   &   \cr
          1   &   &227    &     &972    &     &227    &    &1.  \cr
      }
}
This result is consistent with the Hodge number constraint for 
Calabi-Yau fourfolds 
\eqn\hodgecheck{
44+4h^{(1,1)}+4h^{(3,1)}-2h^{(2,1)}-h^{(2,2)}=0 }
pointed out in \rsvw.

 Starting from any of the configurations \ordcicys\ one simply applies 
the $\IP_1$ split to any projective factor with $n_i>1$ until all 
corresponding $d_a^i=1$ at which point the $\IP_{n_i}$ is 
contracted. 
 
As in the case of threefolds \rcdls\ we also encounter the 
phenomenon of ineffective splitting in those situations 
when the determinantal variety is actually smooth. This can happen 
even though generically we have a higher dimensional singular set. 
An example which illustrates this is given by 
\eqn\ineffsplit{
\matrix{\IP_1\cr \IP_2\cr \IP_3\cr} 
\left[\matrix{2&0\cr 2&1\cr 0&4\cr}\right] ~~\lolra~~
\matrix{\IP_1\cr \IP_1\cr \IP_2\cr \IP_3\cr}
\left[\matrix{0&1&1\cr 0&1&1\cr 1&1&1\cr 4&0&0\cr}\right],} 
for which the relevant set is 
\eqn\singset{
\matrix{\IP_1\cr \IP_2\cr \IP_3\cr}
\left[\matrix{0&1&1&1&1\cr 1&1&1&1&1\cr 4&0&0&0&0\cr}\right]
= \emptyset.}

A complicating feature of splitting transitions between fourfolds 
however is that in contradistinction to threefold splits there exists 
the  possibility of nontrivial splits, or contractions, which connect 
manifold with the same Euler number. Whereas in the case 
of threefolds a split at constant Euler number is necessarily ineffective, 
providing different configurations of the same underlying manifold, it is 
clear from \eulrel\ that nontrivial fourfold splits can occur 
when the singular set consists of a 
configuration of tori. An example of such a transition at constant 
Euler number $\chi=396$ is given by 
\eqn\consteul{
\matrix{\IP_2\cr \IP_4\cr}\left[\matrix{3&0\cr 1&4\cr}\right]_{396}  
~~\lolra ~~~
\matrix{\IP_1\cr \IP_2\cr \IP_4\cr}
    \left[\matrix{0&1&1\cr 3&0&0\cr 1&1&3\cr}\right]_{396}. ~~, 
}
Here the determinantal variety degenerates at the configuration 
\eqn\ninetori{
\matrix{\IP_2\cr \IP_4\cr}\left[\matrix{3&0&0&0&0\cr 
                                        1&1&1&3&3\cr}\right],
}
describing nine tori. 

\subsec{Examples} 
The perhaps simplest example of a splitting transition is the 
split of the sextic 
\eqn\sexsplit{
\IP_5[6]_{2610} ~~\lolra ~~
 \matrix{\IP_1\cr \IP_5\cr}\left[\matrix{1&1\cr 1&5\cr}\right]_{2160}, 
}
where the smooth hypersurface can be defined by the Fermat polynomial 
$$p = \sum_i z_i^6 $$
and a transverse choice of the split configuration is provided 
by 
\eqn\sexcodtwo{\eqalign{
p_1 = & x_1y_1 + x_2y_2 \cr  
p_2 = & x_1\left(y_2^6 + y_4^6 +y_6^6\right) 
        +x_2\left(y_1^6 + y_3^6 + y_5^6\right). \cr
}}
Again the subscripts indicate the Euler numbers, the latter of
which can be obtained by resolving the singular set of
the determinantal variety, given by the genus $g=76$ curve
$\Si=\IP_3[5~~5]$. More precisely the split \sexsplit\ connects
the Hodge diamond of the sextic hypersurface 
\eqn\sexthodge{
\matrix{      &   &       &     &1       &     &      &    &   \cr
              &   &       &0    &        &0    &      &    &   \cr
              &   &0      &     &1       &     &0      &    &   \cr
              &0  &       &0    &        &0    &       &0   &   \cr
          1   &   &426    &     &1752    &     &426    &    &1.  \cr
      }
}
with the Hodge diamond
\eqn\mhodge{
\matrix{      &   &       &     &1       &     &      &    &   \cr
              &   &       &0    &        &0    &      &    &   \cr
              &   &0      &     &2       &     &0      &    &   \cr
              &0  &       &0    &        &0    &       &0   &   \cr
          1   &   &350    &     &1452    &     &350    &    &1.  \cr
      }
}
of the codimension two complete intersection manifold  of \sexsplit.

More interesting splits are obtained by following the strategy described in
the introduction, i.e. by considering fourfolds which are CY$_3$ fibered.
Transitions of this type can be obtained as follows. 
Consider the weighted threefold splits of \rls\  
\eqn\threesplit{
\IP_{(k_1,k_1,k_2,k_3,k_4)}[d]~~\lolra ~~
\matrix{\IP_{(1,1)} \hfill \cr \IP_{(k_1,k_1,k_2,k_3,k_4)}\cr} 
\left[\matrix{1&1\cr k_1&(d-k_1)\cr}\right] 
}
with $d=2k_1\narrowplus k_2\narrowplus k_3\narrowplus k_4$. These 
threefolds 
can be used to construct CY$_3$--fibered fourfolds via the twist map 
\doubref\rhls\rbs. The generic fiber of such manifolds 
then is a quasismooth Calabi-Yau threefold. 
Let $\ell=d/k_4 \in 2\IN +1$. For the hypersurfaces of \threesplit\ this 
amounts to choosing the curve $\cC_{\ell} = \IP_{(2,1,1)}[2\ell]$ and 
applying the twist map 
\eqn\hypetwist{
\IP_{(2,1,1)}[2\ell] ~\times~ \IP_{(k_1,k_1,k_2,k_3,k_4)}[d]  
 ~~\lra ~~ \IP_{(2k_1,2k_1,2k_2,2k_3,k_4,k_4)}[2d] 
}
defined as 
\eqn\hypemap{
\left((x_1,x_2,x_3),(y_1,y_2,y_3,y_4,y_5)\right) \mapsto 
\left(y_1,y_2,y_3,y_4,x_2\sqrt{y_5\over x_1}, x_3\sqrt{y_5\over x_1}\right).
}

For the codimension two threefold in \threesplit\ the twist map produces 
the complete intersection fourfolds 
\eqn\codimtwotwist{\eqalign{
\IP_{(2,1,1)}[2\ell] ~\times ~&  
\matrix{\IP_{(1,1)} \hfill \cr \IP_{(k_1,k_1,k_2,k_3,k_4)}\cr}
\left[\matrix{1&1\cr k_1&(d-k_1)\cr}\right] \cr 
 &~~~~~~~~\lra ~~
\matrix{\IP_{(1,1)} \hfill \cr \IP_{(2k_1,2k_1,2k_2,2k_3,k_4,k_4)}\cr}
\left[\matrix{1&1\cr 2k_1&2(d-k_1)\cr}\right]. \cr  
}}
From this we see that the twist map applied to threefolds which are 
connected via conifold transitions induces splitting transitions 
between fibered fourfolds 
\eqn\foursplit{
\IP_{(2k_1,2k_1,2k_2,2k_3,k_4,k_4)}[2d]~~\lolra ~~
\matrix{\IP_{(1,1)} \hfill \cr \IP_{(2k_1,2k_1,2k_2,2k_3,k_4,k_4)}\cr}
\left[\matrix{1&1\cr 2k_1&2(d-k_1)\cr}\right].
}

Of special interest in this context are fibrations for which the generic
threefold fiber is itself a
K3-fibration\footnote{$^{\ddagger}$}{Several lists identifying such
      examples among the class of hypersurfaces \rweighted\ have been
     described in \rklm\rls\rhly\rakms. Reference \rakms\ also contains 
     a discussion of the much larger class of K3 fibrations described 
    by hypersurfaces in toric varieties.}
whose generic fiber in turn is an elliptic fibration.
Such fourfolds thus are particularly simple elliptic fibrations which are of
use in F-theory. An example is given by the weighted split
\eqn\weightsplit{
\IP_{(8,8,4,2,1,1)}[24] ~\lolra ~
\matrix{\IP_{(1,1)}\hfill \cr \IP_{(8,8,4,2,1,1)}\cr}
\left[\matrix{1&1\cr 8&16\cr}\right],
}
where the lhs manifold is defined by the zero locus of the polynomial  
\eqn\hyps{
p=z_0^3+z_1^3+z_2^6+z_3^{12}+z_4^{24}+z_5^{24} =0
}
and the rhs by the equations
\eqn\codimtwo{\eqalign{
p_1 &= x_1y_1 + x_2y_2  \cr
p_2 &= x_1(y_2^2 + y_4^8 + y_6^{16})
        + x_2(y_1^2 + y_3^4 + y_5^{16}). \cr
}}
The determinantal variety
\eqn\weightdet{
p_{det} = y_1(y_1^2 + y_3^4 + y_5^{16}) -
     y_2(y_2^2 + y_4^8 + y_6^{16}) 
}
is singular on the locus $\Si = \IP_{(4,2,1,1)}[16~16]$, describing a 
smooth curve of genus $g=385$.

The fibration type of the hypersurface of \weightsplit\ 
has been discussed in 
\rbs, where also the Hodge diamond was determined to be 
$(h^{(1,1)}=6,h^{(2,1)}=1,h^{(3,1)}=803, h^{(2,2)}= 3278)$. 
Both manifolds of this split have a nested fibration structure
in which the Calabi-Yau fourfold CY$_4$ is a CY$_3$ fibration 
with threefolds which in turn
are K3 fibrations with elliptic K3 fibers. This iterative 
fibration structure can be summarized in the diagram 
\eqn\nestedfibs{
\matrix{{\rm T}^2 &\lra &{\rm K3} &\lra &{\rm CY}_3 &\lra &{\rm CY}_4\cr
                  &     &\downarrow &   &\downarrow &     &\downarrow \cr
                  &     &\IP_1    &     &\IP_1      &     &\IP_1.  \cr}
}
Using the twist map constructions described in \doubref\rhls\rbs\  
one finds the embedding structure for the hypersurface to be given by 

\eqn\embhyp{
\IP_2[3] \lra \IP_{(2,2,1,1)}[6] \lra \IP_{(4,4,2,1,1)}[12] \lra
\IP_{(8,8,4,2,1,1)}[24],
}
whereas the codimension two space leads to the iterative structure 

\eqn\embwci{
\matrix{\IP_1\cr \IP_2\cr}\left[\matrix{1&1\cr 1&2\cr}\right]
\lra
\matrix{\IP_{(1,1)}\hfill \cr \IP_{(2,2,1,1)}\cr}
        \left[\matrix{1&1\cr 2&4\cr}\right]
\lra
\matrix{\IP_{(1,1)}\hfill \cr \IP_{(4,4,2,1,1)}\cr}
        \left[\matrix{1&1\cr 4&8\cr}\right]
\lra
\matrix{\IP_{(1,1)}\hfill \cr \IP_{(8,8,4,2,1,1)}\cr}
        \left[\matrix{1&1\cr 8&16\cr}\right].
}

This shows that the generic Calabi-Yau threefold fiber of the 
codimension two split is obtained from the hypersurface 
threefold fiber via the split 
\eqn\fibsplit{
\IP_{(4,4,1,1,2)}[12]^{(5,101)} ~\lolra ~
\matrix{\IP_{(1,1)}\hfill \cr \IP_{(4,4,1,1,2)}\cr}
\left[\matrix{1&1\cr 4&8\cr}\right]^{(6,70)},
}
where the hypersurface on the lhs is defined by the polynomial
$$p=z_0^3+z_1^3+z_2^{12}+z_3^{12}+z_4^6 $$
and the codimension two variety on the rhs is defined by
\eqn\codimtwofib{\eqalign{
p_1 =&x_1y_1 + x_2y_2 \cr
p_2 =& x_1(y_2^2 + y_4^8 + y_5^4)
         + x_2(y_1^2 + y_3^8 + y_5^4).\cr
}}
The determinantal threefold 
\eqn\detthree{
\IP_{(4,4,1,1,2)}[12] \ni X^{\sharp} =
\left\{p_{det} 
 = y_1^3-y_2^3+(y_1y_3^8-y_2y_4^8) +(y_1-y_2)y_5^4 =0 \right\} 
}
is singular at the $\IP_{(1,1,2)}[8~8] = 32$ points.
Thus we see that the fiber degenerates at a number of points on the 
curve $\Si$  precisely when the determinantal fourfold 
variety degnerates at $\Si$. 

\newsec{Discriminantal Splitting}
\noindent
There are other simple types of transitions between fourfolds for 
which the global issues mentioned in the previous Section are under 
under control as well. 

Consider the following class of configurations
\eqn\dissplit{
X_{\rm split}~=~\matrix{\IP_1\cr Y\cr} 
\left[\matrix{2&0\cr u&M\cr}\right] 
}
defined by the zero locus of the polynomials 
\eqn\dispollies{\eqalign{
p_1(x_i,y_k) &= \sum_{ij} R_{ij}(y_k) x_ix_j \cr 
p_a(y_k) &=  0,~~~a=2,...,N \cr
}} 
where we again denote the coordinates of the ambient space 
$\IP_1\times Y$ by ($x_i, y_j$). Adapting the threefold analysis 
of \rgh\ to fourfolds shows that this space can be understood as the 
double cover of the space $Y[M]$ branched over a threefold 
$B \subset Y[M]$ except over the singular locus of this threefold. 

More precisely, consider the discriminantal hypersurface in $Y[M]$ defined 
by the polynomial 
\eqn\discr{
p_{\rm dis} = \sum R_{ij}(y)R_{kl}(y) \e^{ik}\e^{jl} 
}
and let $\pi: X_{\rm split} \lra Y[M]$ be the projection of $X_{\rm split}$ 
along the projective curve $\IP_1$. 
For each of the points $y\in Y$ the inverse image $\pi^{-1}(y)$ then 
consists of 
\item{1.} two points if $p_{\rm dis}\neq 0$ 
\item{2.} one point if  $p_{\rm dis}=0$ but at least one of the $R_{ij}$ is 
     non-vanishing 
\item{3.} a copy of $\IP_1$ if all $R_{ij}$s are identically zero.

\noindent
This shows that \dissplit\ is a double cover except for the vanishing 
locus of the discriminant 
\eqn\discrlocus{
B=\{p_{\rm dis} =0\} \subset Y[M],
}
which describes a hypersurface of degree $2u$ in $Y[M]$. This 
discriminant locus 
is singular at the vanishing locus of all the $R_{ij}$s  
\eqn\discsings{
\Si = Y[M~~u~~u~~u],
}
describing a curve (configuration) in $B$. 
Smoothing out this singularity by deforming the discriminant 
then provides an alternative way of resolving the singularity. 
Similar to the situation encountered in the determinantal splitting and 
contraction transitions we can resolve the singular curve of the fourfold 
in two different ways. This then provides a second type of fourfold 
transition. 

A particularly simple class is given by the discriminantal splits
\eqn\siclass{
X~=~\IP_{n+1}[2~~u_2~~\cdots~~u_{n-3}] ~~\lolra~~
\matrix{\IP_1\cr \IP_n\cr}\left[\matrix{2&0&\cdots &0\cr 
                                1&u_2&\cdots &u_{n-3}\cr}\right] 
~=~X_{\rm split}  
}
for which
\eqn\eulreltwo{
\chi(X_{\rm split})- \chi(X) = 6\prod_{a=2}^{n-3} u_a, 
}
leading to a discriminant locus which is singular at 
the curve $\Si = \IP_{n-3}[d_2~\cdots ~d_{n-3}]$. A concrete split 
of this type is described by 
\eqn\displ{
\IP_6[5~~2] ~~\lolra ~~\matrix{\IP_1\cr \IP_5\cr}\left[\matrix{2&0\cr
                                1&5\cr}\right],
} 
with $\Si=\IP_2[5]$, a curve of genus $g(\Si)=6$. 

Even simpler are splits of discriminantal splits of the type 
\eqn\sidis{
\matrix{\IP_1\hfill \cr \IP_{n+1}\cr}
        \left[\matrix{2\cr n+1\cr}\right]
~~\lolra ~~ \IP_{(1,...,n+1)}[2(n+1)]
}
where the hypersurface on the rhs lives in a weighted 
$(n\narrowplus 1)$--space.
 
More interesting however is that by adapting to fourfolds 
certain threefold isomorphisms constructed via fractional 
transformations discussed in \doubref\rlsone\rls\ 
we can construct discriminantal transitions between 
weighted hypersurfaces, such as the split 
\eqn\hypsusplit{
\IP_{(1,1,2,2,2,4)}[12]_{2592} ~~\lolra ~~ \IP_5[6]_{2610}.
}
In order to see this one first notes that we can rewrite the 
weighted hypersurface in \hypsusplit\ as 
\eqn\fiso{
\IP_{(1,1,2,2,2,4)}[12] ~~\sim ~~
\matrix{\IP_{(1,1)}\hfill \cr \IP_{(1,1,1,1,1,2)}\cr}
\left[\matrix{2&0\cr 1&6\cr}\right].
}
This follows most easily by going to the Landau-Ginzburg 
phase in which the addition of trivial mass terms is irrelevant. 
Thus we can equivalently consider the Fermat potential in the 
configuration 
\eqn\lgphase{
\IC_{(1,1,6,6,2,2,2,4)}[12]~ \ni ~ 
\left\{ \sum_{i=1}^2\left(x_i^{12}+y_i^2\right) 
     + \sum_{j=3}^5x_j^6+x_6^3=0\right\}
}
at central charge $c=12$. Here we have denoted the coordinates 
in the weighted complex space by $(x_1,y_1,x_2,y_2,x_3,...,x_6)$. 
Modding out two trivial $\ZZ_2$s and applying the corresponding 
fractional transformation, as explained in \doubref\rlsone\rls, 
we find a third representation of this theory provided by 
\eqn\lgisom{
\IC_{(1,6,1,6,2,2,2,4)}[12]{\Big /} 
\ZZ_2^2\left[\matrix{1&1&0&0&0&\cdots &0\cr
                     0&0&1&1&0&\cdots &0\cr}\right]
~\sim ~  \IC_{(2,5,2,5,2,2,2,4)}[12]
}
with the fractional tansform described by the potential 
\eqn\lgaltpot{
W=\sum_{i=1}^2\left(x_i^6+x_iy_i^2\right) 
       + \sum_{j=3}^5x_j^6+x_6^3.
}
 The manifold phase of this 
Landau-Ginzburg theory can finally be seen to be described 
by the codimension two configuration of \fiso\ by using the 
construction of \rrs.

Repeating now the analysis above we find that this codimension 
two complete intersection manifold is the double cover of 
$\IP_{(1,1,1,1,1,2)}[6]$ branched over the discriminant locus 
described by the threefold $B=\IP_{(1,1,1,1,1,2)}[6~~2]$, which 
is singular at the smooth genus four curve $\Si=\IP_{(1,1,2)}[6]$. 
Deforming the discriminant locus then leads to a variety which 
is isomorphic to the smooth sextic fourfold. 
 
\newsec{Superpotentials}
\noindent 
Different types of Calabi-Yau fourfolds lend themselves for the
compactification of various higher dimensional theories.
If the fourfold $X$ admits an elliptic fibration
$$\matrix{{\rm T}^2 &\lra &X \cr 
                    &     &\downarrow \cr 
                    &     &\cB \cr 
 }$$
with fiber T$^2$ and a threefold base $\cB$, then $M$-theory 
on $X$ leads to type IIB string theory on the base
\doubref\rcvone\rmv. If the base $\cB$ in turn is fibered over a 
surface $\cS$ with the generic
fiber being a sphere $\IP_1$, i.e. we have the structure 
$$\matrix{{\rm T}^2 &\lra   &X \cr
                    &       &\downarrow \cr
          \IP_1     &\lra   &\cB \cr
                    &       &\downarrow \cr 
                    &       &\cS, \cr 
 }$$
then IIB($\cB$) leads to the heterotic
string compactified on an elliptically fibered Calabi-Yau threefold
over $\cS$.

According to the results of ref. \rew\ a superpotential in M-theory
compactification on Calabi-Yau fourfolds is generated by
five-branes wrapping around complex divisors $\cD \subset X$ such
that
\eqn\divholeul{
\chi(\cD, \cO_{\cD}) =1.
}
A sufficient criterion for this to hold clearly is that the
divisors contain no nontrivial holomorphic forms, $h^{(i,0)}=0, i>0$.

It follows from the structure of the second Chern class
that algebraic divisors $\cD \subset X$, described by polynomials
in these manifolds, cannot generate a superpotential because
for the holomorphic Euler number 
\eqn\holeul{
\chi(X, \cL) = \sum_i (-1)^i~{\rm dim} {\rm H}^i(X,\cL) 
}
for any line bundle $\cL$ on a manifold $X$ one computes 
via Hirzebruch-Riemann-Roch 
\eqn\hiriroch{
\chi(\cL) = \int_X ch(\cL)\wedge Td(X) 
}
on a Calabi-Yau fourfold 
\eqn\chidiv{
\chi(\cD, \cO_{\cD}) 
= -{1\over 24} \int c_1(\cD)^2\left(c_1(\cD)^2+c_2(X)\right).
}
Thus for manifolds in which all divisors are of this type, such as 
hypersurfaces in $\prod_i \IP_{n_i}$, no superpotential can be generated.
In \rdgw\ a manifold was described which does contain the requisite
divisors. The manifold can be represented as a complete intersection
of the form
\eqn\edgw{
\matrix{\IP_1\cr \IP_1\cr \IP_2\cr \IP_2\cr}
\left[\matrix{1&1\cr 2&0\cr 3&0\cr 0&3\cr}\right]
}
described by two polynomials whose degrees are described by the columns
of this configuration. The manifolds of this deformation class are
double elliptic fibrations which are also K3 fibered with fibers
which in turn are elliptic.
The relevant divisors identified in \rdgw\ can
be described as blow-ups of $\IP_1\times \IP_2$
along the curve described by the base locus of the K3 fibration.
In the next Section we will describe splitting transitions, and their
inverses, contractions, between the configuration \edgw\ and two
hypersurfaces in which
precisely these divisors originate from small resolutions of the
singular set of the determinantal variety connecting
the smooth manifolds.

\newsec{Generating a superpotential via splitting}
\noindent
Consider the manifold 
\eqn\dgwcontract{
X=\matrix{\IP_1\cr \IP_2\cr \IP_2\cr}
      \left[\matrix{2\cr 3\cr 3\cr}\right].
}
The Euler number of this space can be determined via Cherning
From Lefshetz' hyperplane theorem 
we know that $h^{(1,1)}=3$ and $h^{(2,1)}=0$. Furthermore we 
can determine $h^{(3,1)}=280$ by counting complex deformations. 
Plugging all this into the Euler number leads to the complete 
Hodge half-diamond 
\eqn\mhodge{
\matrix{      &   &       &     &1       &     &      &    &   \cr
              &   &       &0    &        &0    &      &    &   \cr
              &   &0      &     &3       &     &0      &    &   \cr
              &0  &       &0    &        &0    &       &0   &   \cr
          1   &   &280    &     &1176    &     &280    &    &1.  \cr
      }
}
 
It follows from Lefshetz and the Todd formula 
 that manifolds of this type, i.e. hypersurfaces 
embedded in products of ordinary projective spaces, do not lead to 
non-vanishing superpotential. However the manifold above 
 can be split into one that does contain divisors which generate 
a superpotential. More precisely \dgwcontract\  is part 
of a sequence of splits which connects the manifold \edgw, which 
has been shown in \rdgw\ to lead to a superpotential with modular 
properties, with the sextic fourfold 
\eqn\splitone{\eqalign{
&\IP_5[6] ~\lolra ~
    \matrix{\IP_1\cr \IP_5\cr}\left[\matrix{1&1\cr 1&5\cr}\right]
~\lolra ~\matrix{\IP_2\cr \IP_1\cr \IP_5\cr}
     \left[\matrix{0&1&1&1\cr 1&1&0&0\cr 1&1&1&3\cr}\right]~\lolra~
      \cr 
&~~~~~~ \matrix{\IP_2\cr \IP_2\cr \IP_1\cr \IP_5\cr}
   \left[\matrix{0&0&0&1&1&1\cr 0&1&1&1&0&0\cr 1&1&0&0&0&0\cr
         1&1&1&1&1&1\cr}\right]
~\lolra ~ \matrix{\IP_1\cr \IP_2\cr \IP_2\cr}
             \left[\matrix{2\cr 3\cr 3\cr}\right]
~~\lolra ~~  
\matrix{\IP_1\cr \IP_1\cr \IP_2\cr \IP_2\cr}
\left[\matrix{1&1\cr 2&0\cr 3&0\cr 0&3\cr}\right] = X_{\rm split}.}
}
Both of these spaces are elliptic fibrations and the split manifold is 
also a K3-fibration with generic elliptic K3 fibers.

The determinantal hypersurface
\eqn\detone{
\matrix{\IP_1\cr \IP_2\cr \IP_2\cr}
\left[\matrix{2\cr 3\cr 3\cr}\right] \ni 
X^{\sharp} ~=~ \{p_{\rm det} = QT -RS=0\}
}
is singular at the locus 
\eqn\detsind{
\matrix{\IP_1\cr \IP_2\cr \IP_2\cr}
\left[\matrix{2&2&0&0\cr 3&3&0&0\cr 0&0&3&3\cr}\right]= 9\times \Si,
}
where $\Si = \matrix{\IP_1\cr \IP_2\cr}
\left[\matrix{2&2\cr 3&3\cr}\right] $
and $\IP_2[3~~3]=9pts$. The curve 
$\Si$ has Euler number $\chi(\Si) =-54$ and hence is of genus
$g(\Si)=28$. Thus the singular set has 9 different components and 
the splitting formula \eulrel\ becomes 
\eqn\eulex{
\chi(X_{\rm split}) = \chi(X) + 3\cdot 9~\chi(\Si)=288.      
}
We see from this that it is precisely the small resolution of the 
curve $\Si$ which introduces the divisors in $X_{\rm split}$ which are 
responsible for the superpotential. 

The algebraic divisors 
\eqn\mdiv{
X \supset \cD = \matrix{\IP_1\cr \IP_2\cr \IP_2\cr}
\left[\matrix{2&a_1\cr 3&a_2\cr 3&a_3\cr}\right]
}
of the manifold $X$ are transformed by the splitting transition 
into the divisors 
\eqn\splitdiv{
X_{\rm split} \supset
\cD_{\rm split} =
\matrix{\IP_1\cr \IP_1\cr \IP_2\cr \IP_2\cr}
\left[\matrix{1&1&0\cr 2&0&a_1\cr 3&0&a_2\cr 0&3&a_3\cr}\right].
}
On the singular determinantal variety this divisor degenerates into 
a number of points whose resolution  is described by \splitdiv.

A similar discussion applies to the $\IP_1$-split 
\eqn\sexsplit{
X^{'}~=~\matrix{\IP_2\cr \IP_3\cr}
    \left[\matrix{3\cr 4\cr}\right]_{2016} \lolra
\matrix{\IP_1\cr \IP_2\cr \IP_3}
   \left[\matrix{1&1\cr 3&0\cr 0&4\cr}\right]_{288} 
~=~X^{'}_{\rm split}, 
}
which connects the lhs elliptic fibration with Hodge numbers 
$(h^{(1,1)}(X^{'})=2, h^{(2,1)}(X^{'})=0, h^{(3,1)}(X^{'})= 326, 
h^{(2,2)}(X^{'})=1356)$ to the codimension two elliptic fibration of 
the rhs with $\chi(X^{'}_{\rm split})=288$.
The determinantal variety $X^{\sharp '}$ is singular at 
the locus 
\eqn\detsingtwo{
\matrix{\IP_2\cr \IP_3\cr}
\left[\matrix{3&3&0&0\cr 0&0&4&4\cr}\right]= 9\times \Si,
}
where $\Si=\IP_3[4~~4]$ is a genus $g(\Si)=33$ curve, and therefore 
\eulrel\ leads to $\chi(X^{'}_{\rm split})=288$. 

In this example the small resolution of the split transition 
replaces pointwise the curve $\Si \subset \IP_3$ by the projective 
plane, thereby introducing the necessary divisor for a non-vanishing 
superpotential. On the polynomial divisors 
\eqn\divtwo{
X^{'} \supset \cD = \matrix{\IP_2\cr \IP_3\cr}
\left[\matrix{3&a_1\cr 4&a_2\cr}\right], }
which are split into 
\eqn\splitdivtwo{
X^{'}_{\rm split} \supset \cD_{\rm split} =
\matrix{\IP_1\cr \IP_2\cr \IP_3}
\left[\matrix{1&1&0\cr 3&0&a_1\cr 0&4&a_2\cr}\right], }
the small resolution of the curve $\Si$ again translates into the 
resolution of a number of points.  

The manifold $X^{'}_{\rm split}$ can in fact be split and  
contracted at constant Euler number into the split 
manifold \edgw\ via  
\eqn\connect{\eqalign{
\matrix{\IP_1\cr \IP_2\cr \IP_3\cr}
      \left[\matrix{1&1\cr 3&0\cr 0&4\cr}\right]
&~~\lolra ~~
\matrix{\IP_1\cr \IP_1\cr \IP_2\cr \IP_3\cr}
      \left[\matrix{0&1&1\cr 1&1&0\cr 3&0&0\cr 0&1&3\cr}\right]
 ~~\lolra ~~ \cr 
&~~~~~~\matrix{\IP_2\cr \IP_1\cr \IP_1\cr \IP_2\cr \IP_3\cr}
      \left[\matrix{0&0&1&1&1\cr 0&1&1&0&0\cr 1&1&0&0&0\cr 
                    3&0&0&0&0\cr 0&1&1&1&1\cr}\right] 
~~\lolra ~~
\matrix{\IP_2\cr \IP_1\cr \IP_1\cr \IP_2\cr}
    \left[\matrix{0&3\cr 0&2\cr 1&1\cr 3&0\cr}\right]. \cr 
}}
This sequence involves nontrivial determinantal varieties 
which degenerate at configurations of tori, as discussed in 
Section 2.

\vskip .3truein
\noindent
{\bf Acknowledgements}

\noindent
It is a pleasure to thank Gottfried Curio, Dieter L\"ust, 
Werner Nahm, Michael R\"osgen and Andreas Wi\ss kirchen for 
discussions. M.L. thanks the physics department at Bonn University 
and R.S. thanks the physics department at Indiana University, 
South Bend, for hospitality during the course of part of this work.
The work of M.L. was supported in part by a 
Summer Faculty Fellowship.

\listrefs
\bye